# About the quantum mechanical speeding up of classical algorithms

Y.Ozhigov

October 7, 2018

Department of mathematics, Moscow state technological
University "Stankin", Vadkovsky per. 3a, 101472, Moscow, Russia
E-mail: y@oz.msk.ru

## 1 Abstract

This work introduces a relative diffusion transformation (RDT) - a simple unitary transformation which acts in a subspace, localized by an oracle. Such a transformation can not be fulfilled on quantum Turing machines with this oracle in polynomial time in general case.

It is proved, that every function $f : \omega^n \longrightarrow \omega^n$, $\text{card}(\omega) = 4$, computable in time $T(n)$ and space $S(n)$ on classical 1-dimensional cellular automaton, can be computed with certainty in time $O(T^{1/2}S)$ on quantum computer with RDTs over the parts of intermediate products of classical computation. This requires multiprocessor, which consists of $\sqrt{T}$ quantum devices $P_1, P_2, \ldots, P_{\sqrt{T}}$ of $O(S(n))$ size, working in parallel-serial mode and interacting by classical lows. Such a function can be computed only in time $O(S4^{S/2}T/T_1)$ on a conventional quantum computer with oracle for intermediate results of the time $T_1$ of classical computations.

## 2 Introduction

Quantum mechanical computations (QC) are distinct in nature from the classical ones. The point is that a quantum system can be in different classical states simultaneously with the corresponding amplitudes. The vector, composed of these amplitudes completely determines the quantum state of system. The module squared of every amplitude is the probability of detecting the system in the corresponding classical state after observation. Such an observation is the only way to obtain a result of QC. An evolution of such a system is represented by the application of unitary transformation to its vector of amplitudes.

Quantum computers became one of the most popular areas of investigations in theoretical computer science as well as in quantum physics because of that in the past 3-4 years considerable progress has been made in the theory of QC. Since that time when R.Feunmann in the work [Fe] proposed quantum mechanical (Q-M) computer, D.Deutsch in the work [De] gave the first formal model of computations on quantum Turing machine (QTM), and S.Lloyd in the work [Ll] presented the physical scheme of Q-M computational device, the advantages of quantum computations over the classical ones in a variety of particular problems became apparent from the sequence of results (look, for example, at [Be],[Sh], [BE] ). Moreover, A.Berthiaume and G.Brassard in [BB] showed, that Q-M computations even can beat the nondeterministic ones in computations with oracles. But as for absolute (without oracles) computations, advantages of Q-M computers over the probabilistic classical machines were not so obvious.

Situation has been changed in 1994 when P.Shor in his work [Sh] suggested polynomial time quantum algorithms solving problems: of factorization and of finding discrete logarithms. For both problems classical probabilistic algorithms with polynomial time complexity are not known. In Shor's results advantage was taken of discrete Fourier transformation. The following bright result which is closely connected to the present work was obtained by L.Grover in [Gr1]. He was able to construct the quantum algorithm which for the given function $F : \{0,1\}^n \longrightarrow \{0,1\}$ finds such $x$ that $F(x) = 1$, after $O(\sqrt{N})$ quantum evaluations of $F$ provided such $x$ is unique, in opposite to classical probabilistic computers which require $\Omega(N)$ evaluations in average, $N = 2^n$. This is known as the problem of searching. In Grover's algorithm advantage was taken of the so-called Walsh-Hadamard transformation. Later M.Boyer, G.Brassard, P.Hoyer and A.Tapp in [BBHT] extended this method to the case of arbitrary number $t$ of $x$ such that $F(x) = 1$ and provided tight lower bounds for this algorithm depending on $t$. In particular, they showed, that if $t = N/4$ then a solution is found with certainty after single iteration of algorithm! Note, that the general number $O(\sqrt{N})$ of evaluations in quantum searching can not be reduced in view of result of C.Bennett, G.Brassard, E.Bernstein and U.Vazirani [BBBV] who proved, that relative to an oracle chosen at random with probability 1 the class NP cannot be solved in time $o(2^{n/2})$.

Having Grover's algorithm as a good precedent it is interesting to elusidate is it possible to accelerate sufficiently complicated classical algorithms on quantum computers by some regular way of conversion of a classical program to a quantum one. Most likely, this can not be done without some additional information about a classical algorithm. Such an information in its simplest form is an oracle for verifying of the intermediate product. Let a work of classical algorithm on an input word $A$ has the form $x_0(A) \longrightarrow x_1(A) \longrightarrow \ldots \longrightarrow x_T(A)$. An intermediate product of the time $T_1$ is a set $\{\langle A, x_{T_1}(A)\rangle\}$, where $T_1 < T$. A verifier is an oracle for this set.

The following Theorem shows that verifier can accelerate only sufficiently long computations.

**Theorem 1** *Any function $F : \omega^* \longrightarrow \omega^*$ (card$(\omega) = 4$), computable on one dimensional cellular automaton with alphabet $\omega$ in time $T(n)$ and space $S(n)$ can be computed in time $T_q = O(S4^{S/2}T/T_1)$ on a quantum computer with the single processor and verifier for intermediate product of the time $T_1$ for $F$.*

The following Theorem shows the potentials of RDT.

**Theorem 2** *Every function $F(n)$ computable in time $T(n)$ and space $S(n)$ on a classical 1-dimensional cellular automaton can be computed in time $O(T^{1/2}S)$ on a quantum computer with RDT over the parts of intermediate product for $F$, $T_1 = T^{1/2}S$ and multiprocessor.*

Note, that in general case RDT can not be localazed (e.g. represented as a tensor product of small matrix) and can not be replaced by the computations on quantum query machine (look at the work [BBBV] for the definition) of polynomial time complexity (look at the section 7), therefore, generally speaking, the speeding up by Theorem 2 can not be performed on quantum query machine.

## 3  Quantum computer. Multiprocessor

There are three ways to formalize the notion of quantum computation: quantum Turing machines - QTM ( D.Deutsch [De] , E.Bernstein and U.Vazirani [BV] ), quantum cellular automata (J.Watrous [Wa] ), and quantum circuits ( A.Yao [Ya] ). The devices of all these types have the same computational power, as Turing machines. As for the time complexity, the status of quantum computations remains something of enigma, because even with the supposition that P$\neq$NP it is not known, is it possible to solve NP- complete problem on quantum Turing machine with bounded- error probability in polynomial time or not. There are some interesting partial results concerning the relations between the models of quantum computers.

In [BV] E.Bernstein and U.Vazirani proved the relation

$$\text{BPP} \subseteq \text{BQP} \subseteq \text{PSPACE}$$

for the languages decidable in polynomial time on bounded error probabilistic TM (BPP) and with bounded-error probability on QTMs (BQP). A.Yao in [Ya] showed that any QTM can be simulated by the polynomial size quantum Boolean circuit. J.Watrous in [Wa] showed the possibility of simulation of QTMs by 1-dim QCA with linear slowdown .

Here we shall use the model of quantum computer which consists of classical and quantum parts. Quantum part is a system of particles in so-called coherent states, which can change only in accordance with elementary unitary transformations from the given list. The classical part plays a role of controller.

At first describe the computer with the single processor.

A quantum part is a set $\mathcal{E}$ which elements are called cells. $\mathcal{E}$ may be organized as a discrete lattice: $\mathcal{E} = \mathrm{R}^n$ or as a tree with vertices marked by the words: $\varepsilon_{i_1}\varepsilon_{i_2}\ldots\varepsilon_{i_k}$, $k = 0, 1, \ldots; \varepsilon_{i_j} \in \{0, 1\}$, etc. Let $\omega = \{a_1, a_2, \ldots, a_k\}$ be a finite alphabet for the possible states of each cell in $\mathcal{E}$, $k > 1$. An elementary part of quantum computer is referred as qubit. Qubit takes values from the complex 1-dimensional sphere of radius 1: $\{z_0\mathbf{0} + z_1\mathbf{1} \mid z_1, z_2 \in \mathrm{C}, |z_0|^2 + |z_1|^2 = 1\}$. Here $\mathbf{0}$ and $\mathbf{1}$ are referred as pure states of qubit and form the basis of $\mathrm{C}^2$. It is convenient to join some $l$ neighboring qubits in a cell and regard a state of cell as an ensemble of states of all it's qubits, so that these states will be $a_1, a_2, \ldots, a_k$, $k = 2^l$.

A pure state of the quantum part is a function of the form $e : \mathcal{E} \longrightarrow \omega$. If we fix some order on $\mathcal{E} = \{\nu_1, \nu_2, \ldots, \nu_r\}$, a pure state $e$ may be encoded as $|e(\nu_1), e(\nu_2), \ldots, e(\nu_r)\rangle$. We shall identify this state with the word $e(\nu_1)e(\nu_2)\ldots e(\nu_r)$.

Let $e_1, e_2, \ldots$ be all pure states, taken in some fixed order, $\mathcal{H}$ be $k^r$-dimensional Hilbert space with orthonormal basis $e_1, e_2, \ldots, e_N$, $N = k^r$. This Hilbert space can be regarded as tensor product $\mathcal{H}_1 \bigotimes \mathcal{H}_2 \bigotimes \cdots \bigotimes \mathcal{H}_r$ of $k$-dimensional spaces, where $\mathcal{H}_i$ is generated by the possible values of $e(\nu_i)$. A state of quantum part is such element $x \in \mathcal{H}$ that $|x| = 1$. Thus, in contrast to classical devices, quantum device may be not only in pure, but also in coherent states, and this imparts surprising properties to such devices.

An observation of quantum part in state $x = \sum_s \lambda_s e_s$ is a random variable, which takes value $e_s$ with probability $|\lambda|^2$, $s = 1, 2, \ldots$. A pure state $e_s$ is said to be observed in $x$ with probability $|\lambda|^2$. For elements $x = \sum_s \lambda_s e_s$, $y = \sum_s \mu_s e_s \in \mathcal{H}$ their dot product $\sum_s \lambda_s \bar{\mu}_s$ is denoted by $\langle x|y \rangle$, where $\bar{\mu}$ means complex conjugation of $\mu \in \mathrm{C}$, so $\langle x|y \rangle = \overline{\langle y|x \rangle}$.

Any input data for the computer should be represented as initial (pure) state of it: $x_0$.

The essential feature required of quantum mechanical transformations of states in our computer is their unitarity.

The classical part of computer contains a gate array $G$ all inputs of which are in one-to-one correspondence with the cells of $\mathcal{E}$. This gate array consists of elementary gates which belong to the finite set of standard gates with inputs, marked by labels. Some labels are query labels for some fix oracles. At each instant of time computer performs sequentially the following steps:

Step 1. The gate array calculates the function $\Phi_G(v_{i_1}, \ldots, v_{i_p}) = v$ which depends on contents $v_i = e(\nu_i)$ of cells $\nu_{i_1}, \ldots, \nu_{i_p}$ in quantum part and sends the result to the controller. This calculation includes answers of oracles on the cells, corresponding to inputs marked by query labels. If $v = 0$, then controller performs elementary unitary transformation $U$ on the other cells which correspond to inputs marked out by some special labels $l_1, \ldots, l_h$ on condition that there is the fixed number of such inputs, where $U$ depends on these labels (look at Figure 1 ).

Step 2. A gate array $G$ (with labels) changes in accordance with fixed classical rules (for example, as cellular automaton).

After some preliminary calculated number of steps $C(n)$ we stop this process and observe the quantum part to yield the result. Note that the classical part can be in coherent state only in step 1, recovers after this step and changes only in step 2, therefore, it's evolution is determined by the classical lows.

To determine the quantum algorithm the following things should be fixed:

- list of elementary classical gates with labels forming $G : \{G_1, G_2, \ldots, G_s\}$ (where the different $G_i$ can represent the same classical function and differ only in labels ),

- list of elementary unitary transformations $\bar{U} = \{U_1, U_2, \ldots, \}$,

rules for evolution of classical part in step 2,

- function $\{l_1, \ldots, l_h\}^r \longrightarrow \bar{U}$, pointing what transformation should be performed with the marked inputs.

Note that all $U_1, U_2, \ldots$ must be of limited size to obtain a practicle computer. Every transformation $U_j$ has the matrix of size $k^d \times k^d$. The space $\mathcal{E}$ may be divided into $L$ nonintersecting areas called registers, and we denote by $|\bar{x}_1, \bar{x}_2, \ldots\rangle$ such pure state, that $\bar{x}_i$ is placed in $i$-th register (look at [De]). Let $\bar{0}$ denotes the word which conains only letter $0 \in \omega$. Within the framework of the computational model at hand we can perform the transformation $|a, \bar{0}\rangle \longrightarrow |a, F(a)\rangle$ in time $O(T(n))$ for any function $F$, computable in time $T(n)$ on TM or on cellular automaton.

Let's consider again the basis $\mathcal{B} = \{e_1, \ldots, e_N\}$ of Hilbert space $\mathcal{H}$. The equation $\Phi_G(v_{i_1}, \ldots, v_{i_p}) = 0$ selects the set of basis vectors $\mathcal{B}_1$. The transformation from step 1 acts on vector $e \in \mathcal{B}$ as follows: it leaves $e$ unchanged, if $e \notin \mathcal{B}_1$, and in opposite case acts as tensor product $I_1 \bigotimes I_2 \bigotimes \cdots \bigotimes I_{r-d} \bigotimes U$, where $I_m, m = 1, \ldots, r - d$ is identical mapping of $k$-dim space, generated by such $v_{j_m}$ that $j_m$ is not the number of marked input. $U$ acts in $k^d$-dim space, generated by all $v_j$, where $j$ are the numbers of marked inputs. Unitarity of this transformation follows from the unitarity of $U$ and from that $\nu_{i_1}, \ldots, \nu_{i_p}$ are not marked.

Note that Step 1 may be performed also in the following way. Let $\bar{v}$ denotes the values for marked inputs. At first compute the function $\Phi$ conserving the result in the last register: $|v_{i_1}, \ldots, v_{i_p}, \bar{v}, 0\rangle \longrightarrow |v_{i_1}, \ldots, v_{i_p}, \bar{v}, \Phi(v_{i_1}, \ldots, v_{i_p})\rangle$. Then perform the following: $|v_{i_1}, \ldots, v_{i_p}, \bar{v}, v\rangle \longrightarrow |v_{i_1}, \ldots, v_{i_p}, \bar{v}', v\rangle$ where $\bar{v} = \bar{v}'$ if $v \neq 0$, $\bar{v}' = U(\bar{v})$ if $v = 0$ (look at similar transformations in [Sh],[Gr1] ).

A model of computer with multiprocessor differs only in that the space $\mathcal{E}$ is divided into $M$ nonintersecting areas with separated processors working on these areas simultaneously and independently. Their classical parts interact in Step 2 by classical lows. If $\mathcal{E}$ as a whole is in coherent state, then the resulting transformation of all space will be the tensor product: $\tilde{U}_1 \bigotimes \tilde{U}_2 \bigotimes \cdots \bigotimes \tilde{U}_M$ of transformations in all areas.

However, we shall consider a multiprocessor with quantum parts operated independently of one another, e.g. without quantum connections (look at Figure 2).

Leaving aside physical questions concerning quantum devices we now need to focus upon the potentials of the presented multiprocessor's computational circuit.

The work of algorithm on $x_0$ has the form

$$x_0 \longrightarrow x_1 \longrightarrow \cdots \longrightarrow x_N \longrightarrow \cdots.$$

Let $\mathcal{T}$ be the function of the form $\omega^* \longrightarrow \omega^*$, and $\forall A \in \omega^*, i = 1, \ldots, r$ $|\mu_i|^2 \geq 2/3$ iff $e_i = \mathcal{T}(A)$, where $\mu_i = \langle e_i | x_{T(n)} \rangle$, $n = |A|$ (the length of the word $A$). Then the algorithm is said to compute $\mathcal{T}$ with bounded-error probability. To put it another way, given an input word $A$, we imply our iteration $T(n)$ times sequentially, after that observe the quantum part. If the observation gives $e_i$, we conclude, that $\mathcal{T}(x_0) = e_i$. Note that the probability of error $(1/3)$ can be reduced if we iterate this procedure $t$ times and assume the prevailing conclusion. If $|\mu_i| = 1$, then $\mathcal{T}$ is said to be computed with certainty in time $T(n)$.

## 4 Diffusion transform

Every unitary transformation $U : \mathcal{H} \longrightarrow \mathcal{H}$ can be represented by it's matrix $U = (u_{ij})$ where $u_{ij} = \langle U(e_j) | e_i \rangle$ so that for $x = \sum \lambda_p e_p$, $U(x) = \sum \lambda'_p e_p$ we have $\bar{\lambda}' = U\bar{\lambda}$, where $\bar{\lambda}, \bar{\lambda}'$ are columns. In case of only real-valued transformations in place of Hilbert space $\mathcal{H} = \mathtt{C}^N$, Euclidean space $\mathcal{H} = \mathtt{R}^N$ should be regarded. Generally speaking, an idea of efficient quantum algorithms is very simple. Given the input $A$, $x_0(A) = \sum_{p=1}^{N} \mu_p e_p$, we already have the result $e_s = \mathcal{T}(A)$ in this linear combination, but it can be observed only with probability $|\mu_s|^2$ which may be even equal to zero. One multiplication on appropriate unitary matrix $U$ should raise absolute value of amplitude of correct answer $e_s$ on some fairly big constant. Thus after sufficient number of iterations $T(n)$ the required state $e_s$ will have amplitude large enough for observation with probability at least $2/3$. The exact value of $T(n)$ may be of importance, for example in quantum searching the required amplitude grows during $\sqrt{N}$ transformations and after this instant fells down to zero (look at [BBHT]). Now, we'll describe the significant diffusion transform $D$, introduced by L.Grover in [Gr1]. This transform is remarkable for the following reason: been applied to the state $x = \sum_p \mu_p e_p$, $\mu_p \in \mathtt{R}$, it raises by some constant an absolute value of amplitude $\mu_s$, opposite to average amplitude.

Diffusion transform $D$ is defined by it's matrix $D$:

$$d_{ij} = \begin{cases} 2/N, & \text{if } i \neq j, \\ -1 + 2/N, & \text{if } i = j. \end{cases}$$

Unitarity of $D$ can be easily verified. Note that $D = WRW$, where $R$ is the rotation matrix, defined by

$$r_{ij} = \begin{cases} 0, & \text{if } i \neq j, \\ 1, & \text{if } i = j = 0, \\ -1, & \text{if } i = j \neq 0, \end{cases}$$

and $W$ is Walsh - Hadamard transform, defined by $w_{ij} = (-1)^{\bar{i}\bar{j}}/\sqrt{N}$, where $\bar{i}, \bar{j}$ are binary representations of $i, j$, and $\bar{i}\bar{j}$ denotes the bitwise dot product of the two strings. For any state $x = \sum_{p=1}^{N} \lambda_p e_p$ an average amplitude is taken as $x_{av} = \sum_{p=1}^{N} \lambda_p/N$. Hereafter $\mathcal{H}$ denotes (real) Euclidean space.

**Proposition 1** (Grover, [Gr1]). *For every state $x$*

$$\langle e_p|x\rangle - x_{av} = x_{av} - \langle e_p|D(x)\rangle. \tag{1}$$

**Proof**

Observe that $D = 2P - I$, where $I$ is identity matrix and $P = (p_{ij})$, $p_{ij} = 1/N$ for all $i, j$. Then $D(x) = 2x_{av}(e_1 + e_2 + \cdots + e_N) - x$ which yields (1). □

This means that $D$ is the inversion about average. We need this property related to a subspace $\mathcal{H}_0 \subseteq \mathcal{H}$. Let $e_1, \cdots, e_M, \cdots, e_N$ be basis of $\mathcal{H}$, $\mathcal{H}_0$ be subspace of $\mathcal{H}$ with basis $e_1, \cdots, e_M$. Define the relative diffusion transform $D^{\mathcal{H}_0}$ by

$$d_{ij}^{\mathcal{H}_0} = \begin{cases} 2/M, & \text{if } i \neq j; i, j \in \{1, \ldots, N\}, \\ -1 + 2/M, & \text{if } i = j \in \{1, \ldots, M\}, \\ \delta_{ij}, & \text{in other cases.} \end{cases}$$

For the state $x = \sum_{p=1}^{M} \lambda_p e_p$ average amplitude related to $\mathcal{H}_0$ is taken as $x_{av}^{\mathcal{H}_0} = \sum_{p=1}^{M} \lambda_p/M$. Proposition 1 with the proof can be extanded to RDT as follows.

**Proposition 2** *For every $p = 1, 2, \cdots, M$*

$$\lambda_p - x_{av}^{\mathcal{H}_0} = x_{av}^{\mathcal{H}_0} - \langle e_p|D^{\mathcal{H}_0}(x)\rangle. \tag{2}$$

Using Proposition 1 Grover in [Gr1] implies difffusion transforms sequentially alternating them with the simple transform which changes the sign of the target state and thus increases the amplitude of target state ( initially taken as $1/\sqrt{N}$ ) approximately on $1/\sqrt{N}$ in each iteration. Therefore his algorithm takes $O(\sqrt{N})$ steps to make the target state really observed.

More exactly, let $x_0 = (e_1 + e_2 + \cdots + e_N)/\sqrt{N}$ be initial state which can be simply prepared by applying Walsh - Hadamard transform to the pure state $e_1$. Let $e_N$ be the target state. One step of Grover's algorithm consists of sequential implementation of $e_N$ -rotation

$$R^N(e_i) = \begin{cases} e_i, & i \neq N, \\ -e_N, & i = N \end{cases}$$

and following diffusion transform $D$. Proposition 1 implies that the amplitude of target atate $e_N$ grows approximately on $1/\sqrt{N}$ as a result of every step for some fixed general number $\Omega(\sqrt{N})$ of steps.

Walsh-Hadamard transform can be represented as a tensor product of $n$ matrices $\begin{pmatrix} 1/\sqrt{2} & 1/\sqrt{2} \\ 1/\sqrt{2} & -1/\sqrt{2} \end{pmatrix}$ therefore the diffusion transform can be localized and Grover's algorithm can be fulfilled on a quantum Turing machine with an oracle for the target state. In the section 7 it is shown that RDT in the general case can not be fulfilled in a polynomial on $n$ time on a quantum computer with oracles for $\mathcal{H}_j$.

## 5 Q-M speeding up

Suppose, we have a function $F: \omega^* \longrightarrow \omega^*$ computable in time $T(n)$ and in space $S(n)$ on classical Turing machine or cellular automaton. A consideration of the case $T > n^2$ would suffice to prove Theorem 1. Without loss of generality here we can assume that the input data for this automaton include all space required for computations of $F$, e.g. $S(n) = n$. Our goal is to compute $F$ faster then $\Omega(T(n))$ on quantum computer with RDTs and multiprocessor. On classical computers only some peculiar problems may be solved faster with multiprocessors, for example, the problem of searching. We are going to describe the general method to perform all sufficiently long computations on quantum computers with RTDs and multiprocessors faster than on classical ones.

Let $f : \omega^* \longrightarrow \omega^*$ denote one step in the work of classical algorithm $F$. In case $F$ is 1-dim cellular automaton with radius $R$ the neighborhood of radius $R$ of each $i$-th letter in $\bar{a}$ determines $i$-th letter in $f(\bar{a})$. Without loss of generality we can assume $\text{card}(\omega) = 4$, because any cellular automaton can be simulated without slowdown by such cellular automaton of appropriate radius. Define inductively for every $a \in \omega^*$ $f^{(0)}(a) = a$, $f^{(m)}(a) = f(f^{(m-1)}(a))$, so $f^{(m)}$ is $m$-iteration of $f$, $f^{(C)} = F$.

At first we prove Theorem 1.

**Proof of Theorem 1**

We can assume that $T > n4^{n/2}$ because otherwise $T = O(T_q)$. Prepare the state $\frac{1}{4^{n/2}} \sum_x |x\rangle$. Now, using an oracle and applying Grover's algorithm we obtain $f^{(T_1)}(x_0)$ in time $4^{n/2}n$. Then iterate this procedure and obtain sequentially $f^{(2T_1)}(x_0), f^{(3T_1)}(x_0), \ldots, f^{(T)}(x_0)$, which takes $T/T_1$ iterations, and we obtain the result in time $T_q$. Theorem 1 is proved.

**Proof of Theorem 2**

Fix $n$ and integers $T_1, T_2 : T_1 T_2 = T(n)$. Let $T_1$ independent processors be given: $P_1, P_2, \ldots, P_{T_1}$, every $P_i$ with quantum part $B_i = \{1, 2, \ldots, 3n\}$. The pure states of all $P_i$ will have the form: $|a_1, \ldots, a_{3n}\rangle$, where all $a_i \in \omega$.

At first prepare the state
$$\frac{1}{k^{n/2}} \sum_{\bar{a}} |\bar{0}, \bar{a}, \bar{0}\rangle,$$
in each processor, where $\bar{0} = 0^n$, $\bar{a} = (a_{i_1}, \ldots, a_{i_n})$, which can be done by the simultaneous implying Walsh - Hadamard transformation to the states of the form $|\bar{0}, 0, \ldots, 0, a, 0, \ldots, 0, \bar{0}\rangle$. Then we calculate $T_2$-iteration of $f$ in the last registers to yield the state
$$X_0 = \frac{1}{k^{n/2}} \sum_{\bar{a}} |\bar{0}, \bar{a}, f^{(T_2)}(\bar{a})\rangle$$
for all processors. This takes $O(T_2)$ steps.

After that the processors work in serial mode computing sequentially the intermediate results $\text{tar}_1, \text{tar}_2, \ldots, \text{tar}_{T_1}$, where
$$\text{tar}_i = \frac{1}{k^{n/2}} \sum_{\bar{a}} |f^{(iT_2)}(x_0), \bar{a}, f^{(T_2)}(\bar{a})\rangle,$$
where $x_0$ is some fixed input word for $F$ of the length $n$.

Beginning with $\text{tar}_i$ the processor $P_i$ achives the pure state
$$\text{tar}^*_{i+1} = |f^{(iT_2)}(x_0), f^{(iT_2)}(x_0), f^{(iT_2+T_2)}(x_0)\rangle$$
in time $O(n^2)$ and then prepares the state $\text{tar}_{i+1}$ of the following processor $P_{i+1}$ which is initially set to the state $X_0$. The last passage is quite clear, it takes one instant of time, and the point is to describe the first passage: $\text{tar}_i \longrightarrow \text{tar}^*_{i+1}$.

We omit indices, now $\text{tar}^*$ is our target state. Let $\mathcal{H}_0$ be Euclidean space with orthonormal basis $\mathcal{B}_0$ consisting of all vectors of the form
$$|f^{(iT_2)}(x_0), \bar{a}, f^{(T_2)}(\bar{a})\rangle. \tag{3}$$
We have: $\text{tar}^* = |\alpha_1, \ldots, \alpha_n, \alpha_1, \ldots, \alpha_n, \beta_1, \ldots, \beta_n\rangle$. $\mathcal{B}_s$ denotes the set of all vectors of the form $|\alpha_1, \ldots, \alpha_n, \alpha_1, \ldots, \alpha_s, \gamma_{s+1}, \ldots, \gamma_{2n-s}\rangle$ from $\mathcal{B}_0$, $s = 0, \ldots, n$. Define $\mathcal{H}_s$ as subspace of $\mathcal{H}_0$ spanned by all vectors from $\mathcal{B}_s$. Then
$$\dim \mathcal{H}_s = k^{n-s}; \{\text{tar}\} = \mathcal{H}_n \subset \mathcal{H}_{n-1} \subset \cdots \subset \mathcal{H}_1 \subset \mathcal{H}_0.$$

Now apply sequentially, for $j = 1, 2, \ldots, n$ the following procedure, beginning with tar.

a) Rotation of all $\xi \in \mathcal{B}_j$.
b) Following RDT $D^{\mathcal{H}_{j-1}}$.

After that we observe the quantum part. If $k = 4$, then in the instant of observation "tar*" has amplitude 1. To show this we need the following Lemma. Let $\chi_j$ be the result of step $j$ of our procedure a), b), $\chi_0 = \text{tar}$.

**Lemma 1** *For all* $\xi \in \mathcal{B}_j, \ j = 0, 1, \ldots, n$

$$\langle \chi_j | \xi \rangle = (3 - \frac{4}{k})^j / k^{\frac{n}{2}}.$$

*Proof*

Induction on $j$. Basis follows from the choise of $\chi_0$. Step. Let $j > 0$. By inductive hypothesis for all $\xi \in \mathcal{B}_{j-1}$ $\langle \chi_{j-1} | \xi \rangle = (3 - 4/k)^{j-1}/k^{n/2}$. Denote this by $R$. After rotation amplitude of all $\xi \in \mathcal{B}_j$ is $-R$, and therefore, average amplitude before diffusion is $((k^{n-j+1} - k^{n-j})R - k^{n-j}R)/k^{n-j+1} = ((k-1)R - R)/k$. By Proposition 2 for $\xi \in \mathcal{B}_j$

$$\langle \chi_j | \xi \rangle = R + 2(kR - 2R)/k = R(1 + 2 - 4/k) =$$
$$= (3 - 4/k)^{j-1}(3 - 4/k)/k^{n/2} = (3 - 4/k)^j/k^{n/2}.$$

Lemma 1 is proved. Now put $k = 4$, Lemma 1 yields $\langle \chi_j | \xi \rangle = 2^j/k^{n/2}$. Consequently, $\langle \chi_n | \text{tar}^* \rangle = 1$.

This procedure, increasing amplitude of the target state consists of $n$ implementations of diffusion and rotation transformations each of which takes $O(n)$ time. Consequently, the general time of computation of the function $F$ on the quantum computer with RDTs $D^{\mathcal{H}_{j-1}}$ is $\mathcal{T}(n) = O(\frac{T}{T_1} + n^2)$.

The minimal time for $T = \text{const}$ will be if

$$-\frac{aT}{T_1^2} + 2bn^2 = 0,$$

for some constants $a, b$, which yields $T_1 = O(T^{1/2}/n)$. For this choise of $T_1$ we have $\mathcal{T}(n) = O(T^{1/2}n)$. Theorem 2 is proved.

## 6 The power of RDT

**Theorem 3** *1) There is no regular way to fulfill all RDTs with the corresponding oracles on a quantum Turing machine in polynomial time.*
*2) Any device which is able to perform RDTs over the sets, localized by arbitrary oracles also can find the solution of equation $f(x) = 1$ for a given oracle for $f$ in polynomial time with high probability provided this solution is unique.*

**Proof of Theorem 3**

**Lemma 2** *Given the state*

$$x_0(f) = \frac{1}{k^{n/2}} \sum_{\bar{b}} |f(\bar{b}), \bar{b}\rangle$$

*and the value $\bar{a}$ where $f$ is unknown one-to-one function $\omega^n \longrightarrow \omega^n$, $k = \text{card}(\omega) = 4$, it is possible to find $f(\bar{a})$ with sertainty in time $O(n^2)$ on a computer with RDTs over the sets of the form*

$$N_{e_1 \ldots e_k} = \{\langle f(\bar{a}), \bar{a} \rangle | \exists e_{k+1}, \ldots, e_n : \ \bar{a} = e_1 \ldots e_k e_{k+1} \ldots e_n\}.$$

*Proof*

Let $\mathcal{B}_j$ be the set of such vectors of the form $|f(\bar{b}), \bar{b}\rangle$, that the first $j$ components of $\bar{a}$ and $\bar{b}$ are equal, $\mathcal{H}_j$ be Euclidean space with the basis $\mathcal{B}_j$.

Then $\{|f(\bar{a}), \bar{a}\rangle\} = \mathcal{H}_n \subset \mathcal{H}_{n-1} \subset \cdots \subset \mathcal{H}_0$. Apply sequentially for $j = 1, 2, \ldots, n$: rotation of all $\xi \in \mathcal{B}_j$ and RDT $D^{\mathcal{H}_{j-1}}$. This results in $|f(\bar{a}), \bar{a}\rangle$ by virtue of Lemma 1. □

1) Consider the computation with RDTs from Lemma 2, depending on $f$: $x_0(f) \longrightarrow x_1(f) \longrightarrow \ldots \longrightarrow x_p(f)$, $p = O(n^2)$.

Consider some other function $\tilde{f}$ which differs from $f$ only on two arguments, one of which is $\bar{a}$. Then $\|x_0(\tilde{f}) - x_0(f)\| \leq 2/\sqrt{N}$, $N = 4^n$. Let us assume that RDTs can be fulfilled in polynomial time with the corresponding oracles for $N_{e_1\ldots e_k}$. By the definition of $\tilde{f}$ the corresponding post query states $x_m$ $m = 1, 2, \ldots, p$ for computations of $f(\bar{a})$ and $\tilde{f}(\bar{a})$ differs on $mP(n)/\sqrt{N}$ where $P(n)$ is a polynomial, as it is shown in the article [BBBV]. But it is impossible, because for the final pure states $x_p(f) = |\bar{a}, f(\bar{a})\rangle$, $x_p(\tilde{f}) = |\bar{a}, \tilde{f}(\bar{a})\rangle$ we have $\|x_p(f) - x_p(\tilde{f})\| = \sqrt{2}$. Point 1) is proved.

2) Let's regard an oracle for the function $f$ and find the solution of equation $f(x) = 1$ with RDTs. Put $\mathcal{B}_m = \{\langle x, y\rangle | \ y = f(x) \& (y = 1 \ \vee \exists x': \ x = (0,0)^m x')\}$, $(0,0) \in \omega$. Then we have: $|\text{card}(\mathcal{B}_m) - 4^{n-m}| \leq 1$, because the solution is unique. Therefore, applying the algorithm from the proof of Lemma 2 we obtain the desired $x$ with high probability in time $O(n^2)$. Theorem 3 is proved.

# 7 Final Notes

It is readily seen that RDT in general case is too powerful to be realized on any possible computer unless there is a major breakthrough in complexity theory. But to accelerate a computation of a particular function all that we need is RDTs corresponding to the intermediate products of classical computation of $f$. It is interesting to find a nontrivial function for which such RDTs can be localized. There are the other interesting ways to speed up computations. Specifically, for the interactive algorithms the method of quantum telecomputations (TC) recently proposed by L.Grover in [Gr2] holds much promise. TC allows to reduce the amount of information transmitted between the different parts of quantum computer and thus may save time for computations.

# 8 Acknowledgments

I am grateful to Peter Hoyer whose remarks helped me to correct some errors in previous version . I am also grateful to Andre Berthiaume, Peter Hoyer and Lov Grover for the helpful information and discussions concerning quantum computations.

# References

[BE]  A.Barenco, A.Ekert, *Quantum Computation*, Acta physica slovaca v.45, No. 3, pp 1-12

[BBBV]  C.H.Bennett, E.Bernstein, G.Brassard, U.Vazirani, *Strenths and Weakness of Quantum Computing*, To appear in SIAM Journal on Computing. On line: http://xxx.lanl.gov/archive/quant-ph/9701001)

[BV]  E.Bernstein, U.Vazirani, *Quantum complexity theory*, Manuscript, ( preliminary version in Proceedings of the 25 Annual ACM Symposium On Theory of Computing, 1993, pp 11-20 ),

[BB]  A.Berthiaume, G.Brassard, *Oracle quantum computing*, Journal of modern optics, 1994, vol.41, NO. 12, pp 2521-2535

[Be]  A.Berthiaume, *Quantum Computation*, Manuscript

[BBHT]  M.Boyer, G.Brassard, P.Hoyer, Alain Tapp, *Tight bounds on quantum searching*, Fourth Workshop on Physics and Computation, Boston University, 22-24 Nov. 1996. On line: http://xxx.lanl.gov/archive/quant-ph/9605034

[De]  D.Deutsch, *Quantum theary, the Church-Turing principle and the universal quantum computer*, Proc.R.Soc.Lond. A 400, pp 97-117 (1985),

[Fe] R.Feynman, *Simulating physics with computers*, Internat.J.Theoret.Phys.,21, pp 467-488

[Gr1] L.K.Grover, *A fast quantum mechanical algorithm for database search*, Proceedings, STOC 1996, Philadelphia PA USA, pp 212-219. On line: http://xxx.lanl.gov/archive/quant-ph/9605043

[Gr2] L.K.Grover *Distributed Quantum Computation*. On line: http://xxx.lanl.gov/archive/quant-ph/9704012

[Ll] S.Lloyd, *A Potentially realizable Quantum Computer* Science, 17 September 1993, v. 261, pp 1569-1571

[Sh] P.W.Shor, *Polynomial-Time Algorithms for Prime Factorization and Discrete Logarithms on Quantum Computer*, On line: http://xxx.lanl.gov/archive/quant-ph/9508027 v2 (A preliminary version in Proceedings of the 35th Annual Symposium on Foundations of Computer Science, Santa Fe, NM, Nov. 20-22, 1994, IEEE Computer Society Press, pp 124-134)

[Wa] J.Watrous *On One-Dimensional Quantum Cellular Automata*, Proceedings of the 36th Annual IEEE Symposium on Foundations of Computer Science, 1995

[Wo] S.Wolfram Cellular Automata and Complexity: Collected Papers Addison-Wesley, 1994

[Ya] A.Yao, *Quantum Circuit Complexity*, Proceedings 34th Annual Symposium on Foundations of Computer Science (FOCS), 1993, pp 352-361